\documentclass[11pt]{article}
\usepackage{times}
\usepackage{geometry}
\usepackage[utf8]{inputenc}
\usepackage{enumitem,amssymb}
\usepackage{ragged2e}
\newlist{thematic}{itemize}{8}
\setlist[thematic]{label=$\square$}
\usepackage{pifont}
\usepackage{fancyhdr}
\usepackage[USenglish]{babel}
\usepackage[utf8]{inputenc}
\usepackage[T1]{fontenc}
\usepackage{epsfig}
\usepackage{wrapfig}
\usepackage{color}
\usepackage{paralist}

\usepackage[vflt]{floatflt}
\usepackage{longtable}
\usepackage{caption}
\usepackage{jheppub}   
\usepackage[dvipsnames,svgnames,x11names]{xcolor}
\usepackage[all]{hypcap} 
\usepackage{amssymb,amsmath}
\usepackage{longtable}
\usepackage{amssymb}
\usepackage{amsmath, xspace}
\usepackage{graphics,graphicx} 
\usepackage{rotating}
\usepackage{color}
\usepackage{xcolor}
\usepackage{multirow}
\usepackage{subfigure}
\usepackage{sectsty}
\usepackage[outercaption]{sidecap}
\sidecaptionvpos{figure}{c}
\usepackage{natbib}

\definecolor{DarkGreen}{rgb}{0.0, 0.3, 0.0}
\definecolor{purple}{rgb}{0.5, 0.0, 0.5}
\definecolor{red}{rgb}{1, 0.0, 0.0}
\definecolor{green}{rgb}{0, 1.0, 0.0}

\def\3he{$^3{\rm He}$}

\def\lsim{\mathrel{\lower2.5pt\vbox{\lineskip=0pt\baselineskip=0pt
           \hbox{$<$}\hbox{$\sim$}}}}

\def\gsim{\mathrel{\lower2.5pt\vbox{\lineskip=0pt\baselineskip=0pt
           \hbox{$>$}\hbox{$\sim$}}}}

\let\oldbibitem\bibitem
\renewcommand{\bibitem}{\item[\textbullet]\oldbibitem}

\begin{document}

\raggedright
\huge
A wide-field, multi-line survey of CO in the Magellanic Clouds at parsec-scale resolution: 
characterising the molecular gas content with a 50-m single-dish submillimeter telescope
\linebreak
\bigskip
\normalsize

\textbf{Authors:} 
Francisca (Ciska) Kemper (ciska.kemper@icrea.cat, ICE-CSIC/ICREA/IEEC, Spain);
Rosie Chen (MPIfR, Germany);
Axel Weiss (MPIfR, Germany);
Caroline Bot (Observatoire astronomique de Strasbourg, France);
Fr\'ed\'eric Galliano (CEA/Saclay, France);
Suzanne Madden (CEA/Saclay, France);
Oscar Morata (ICE-CSIC, Spain);
Naslim Neelamkodan (United Arab Emirates University, UAE);
Rebeca Pirvu (ICE-CSIC, Spain);
Monica Rubio (U. Chile, Chile);
Kazuki Tokuda (Kagawa University, Japan)
\linebreak

\textbf{Science Keywords:} 
ISM and star formation; Nearby galaxies; Molecular gas; Submillimeter spectroscopy; Galaxy evolution; Large-scale surveys
\linebreak

 \captionsetup{labelformat=empty}
\begin{figure}[h]
   \centering
\includegraphics[width=.7\textwidth]{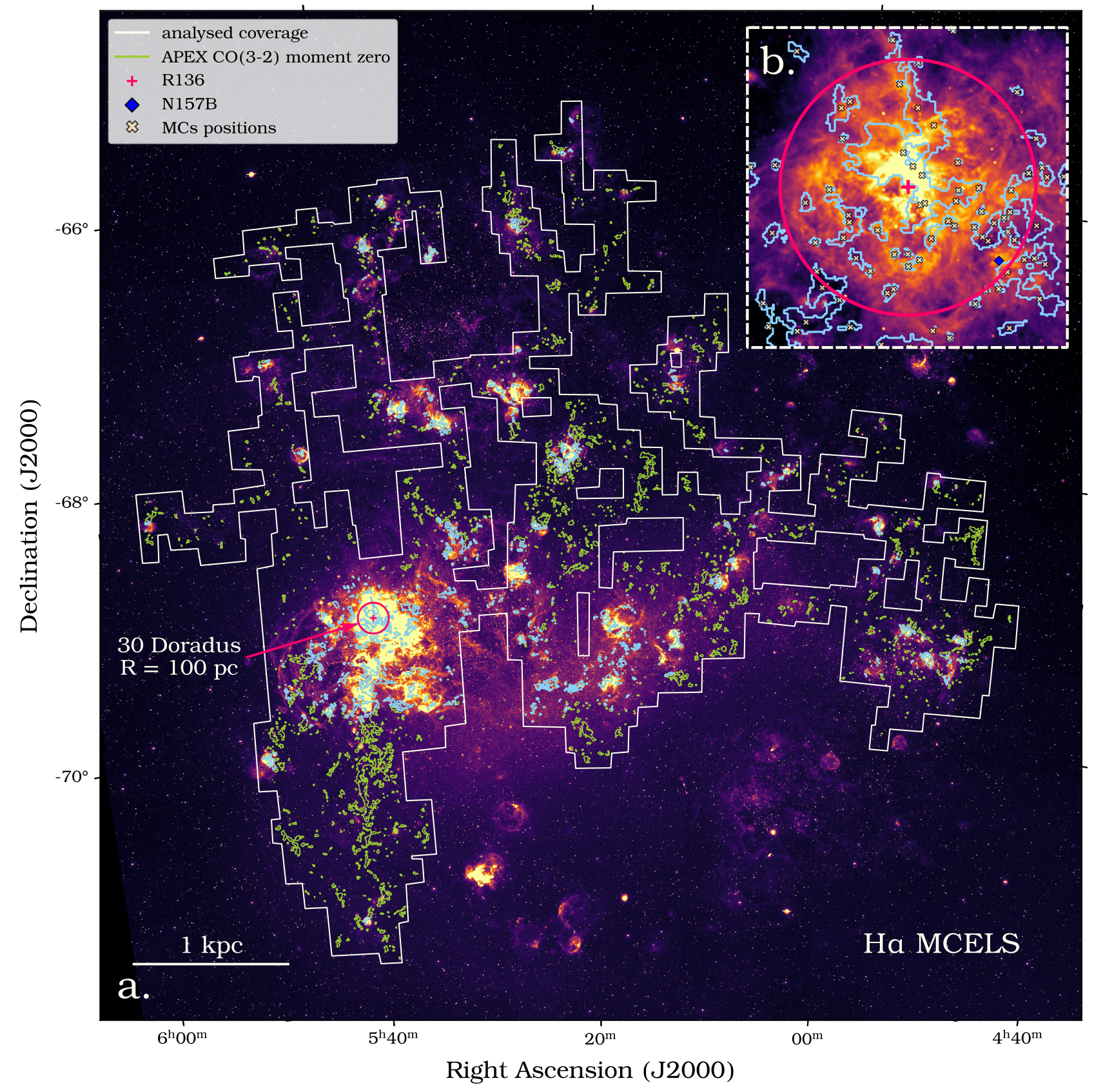}
   \caption{CO(3--2) line intensity map of the Large Magellanic Cloud obtained at 5 to 6 pc resolution with APEX, overlayed on the H$\alpha$ MCELS map \cite{Grishunin_24_Observing,Smith1998}. The white outline represents the surveyed area, and the green contours the detections of CO(3--2).}
\end{figure}
\vspace{-15mm}

\setcounter{figure}{0}
\captionsetup{labelformat=default}


\pagebreak

\paragraph{Abstract}
The Large and Small Magellanic Clouds (LMC, SMC) are nearby dwarf galaxies whose proximity uniquely enables molecular cloud-scale resolution observations across the entire Magellanic system, a capability unmatched in any other external galaxy. Their low metallicities resemble conditions near the peak of cosmic star formation, allowing resolved studies of interstellar medium (ISM) phases, molecular cloud lifecycles, and feedback processes that regulate galaxy evolution. Comprehensive, wide-field spectroscopic mapping of CO and its isotopologues in different transitions, complemented by [C\,I] observations, and combined with already existing HI and HII surveys, will calibrate star-formation laws and gas-phase partition under low-metallicity conditions and furnish benchmarks for interpreting high-redshift galaxies and cosmological simulations. This science requires a large-aperture, wide-field submillimeter single dish with multi-pixel spectroscopic capabilities, operated from a high, dry site such as Chajnantor, to deliver fast, sensitive, high-resolution, degree-scale mapping with total-power fidelity. We present the case for a full molecular atlas of the Magellanic system, the enabling facility requirements, and the transformative impact on galaxy evolution studies.

\vspace*{-12pt}
\section{Scientific context and
motivation}\label{scientific-context-and-motivation}
\vspace*{-6pt}

The Magellanic Clouds are in a privileged position in the extragalactic landscape. At distances of $\sim$50 and $\sim$62 kpc \cite{Pietrzynski_13_Eclipsing,Graczyk_20_Distance}, the LMC and SMC are close enough to resolve individual molecular clouds in different wavelength regimes, yet extended enough on the sky to encompass entire galactic ecosystems in wide-field observations. Their sub-solar metallicities ($Z_{\rm LMC}\approx0.5\,Z_\odot$; $Z_{\rm SMC}\approx0.2\,Z_\odot$) bracket conditions prevalent during the peak of cosmic star formation ($z\sim1$--4), making them analogues for environments where the majority of stars in the universe formed. The LMC hosts 30~Doradus, the most local analogue to a starburst environment, while the ongoing interaction between the LMC and SMC, and with the Milky Way, drives large-scale gas flows (e.g., the Magellanic Bridge and Stream) and strong environmental gradients. These interactions create a natural laboratory for studying how tidal forces, ram pressure, and feedback regulate the conversion of gas into stars, an important driver for galaxy evolution itself. 

The outside vantage point and relatively face-on geometries minimize line-of-sight confusion, enabling spatially coherent studies of feedback, cloud formation, and star formation. The dust distribution and star formation across both the LMC and SMC have been mapped using \emph{Spitzer} \cite{Meixner_06_Spitzer,Gordon_11_Surveying} and \emph{Herschel} \cite{Meixner_13_Herschel}, and deep H\,I imaging with ASKAP \cite{Pingel2022,Dickey_13_GASKAP} and the MCELS and DeMCELS H$\alpha$ surveys \cite{Smith1998,Points_24_Dark} have characterized the atomic and ionized phases. However, for the LMC, a full map of the molecular gas phase, H$_2$, has only been obtained in the tracer CO (1--0) with the 4-m NANTEN antenna at relatively low spatial resolution (2.6'; 38 kpc) and low sensitivity (0.07 K r.m.s.) \cite{Fukui_99_First,Fukui_08_Second}. A similar NANTEN survey was performed for the SMC \cite{Mizuno_01_CO}.  Higher sensitivity and higher resolution follow-up in CO(1--0) was only performed in a patchwork of the brightest regions, e.g. with 11 pc resolution for the LMC \cite{Wong_11_Magellanic}. Targeted ALMA observations add high spatial resolution in selected fields \cite[e.g.]{Wong_22_30}, but their small instantaneous field-of-view and interferometric filtering of extended emission preclude uniform coverage and complete recovery of large-scale structures. Due to its smaller size, it has been more feasible to survey larger sections of the SMC in, e.g. CO(2--1) with APEX at 9 pc resolution \cite{Saldano_23_CO2-1}, and ALMA-ACA at 2 pc resolution \citep{Tokuda_21_Unbiased}. An extensive, but not full coverage, CO(3--2) survey with APEX has revealed comprehensive cloud structure in the LMC at a resolution of 20$''$ (5--6 pc; see Fig. cover page; \cite{Grishunin_24_Observing}) and SMC (\cite{Saldano_24_SuperCAM}; Chen et al. in prep.).
This, however, leaves critical gaps in connecting dust, and atomic and ionized gas, to the structure, kinematics, and excitation of molecular clouds across the metallicity, radiation field, and pressure gradients that occur in the Magellanic system. A full map of a single CO transition, combined with small areas of multiple line transitions at relatively low sensitivity and spatial resolution, limits our understanding of star formation efficiency, feedback, and ISM phase balance in low-metallicity environments.

In their role as analogues of high-redshift galaxies, the Magellanic Clouds are benchmarks for testing ISM physics at low metallicity: the impact of reduced dust-to-gas ratio on shielding, the efficiency of molecule formation, and the coupling between turbulence and star formation. Their well-determined distances allow absolute calibrations of cloud sizes, masses, and energy injection scales. Furthermore, the extensive archival footprints in the infrared, radio, and optical regimes enable multi-wavelength synergies, while the face-on viewing angle simplifies the radiative transfer and dynamical modeling required to interpret observations.
\vspace*{-12pt}
\subsection{The resolution--extent trade-off}
\vspace*{-6pt}
A central reason why the Magellanic Clouds are indispensable for galaxy evolution is the trade-off between angular size and spatial resolution. While southern spirals such as NGC~300, NGC~55, NGC~247, NGC~253 and NGC~4945 are more extended, more massive and more structurally rich, their larger distances (2--4 Mpc) imply that a single-dish beam of 10$''$ corresponds to 100--180 pc, blending entire cloud complexes and erasing substructure. By contrast, the same 10$''$ beam in the LMC/SMC resolves 2.4--3.0 pc, enabling cloud-scale mapping across whole galaxies from an outside vantage point. Even the nearest spiral galaxies, M31 and M33, observable only from the Northern hemisphere, are only a factor of two closer compared to the nearest Southern spirals, with a corresponding improvement in resolution.

Thus, the Magellanic Clouds provide a unique opportunity to carry out a fully sampled, \emph{cloud-scale} spectroscopic survey of the entire molecular reservoir, while maintaining the outside vantage that is impossible in the Milky Way. The resulting demographic studies of cloud masses, sizes, linewidths and lifetimes become statistically robust and directly comparable across gradients in metallicity, radiation field, and pressure. These are indeed the parameters that drive galaxy evolution.

\vspace*{-12pt}
\subsection{Inventory of phases of the ISM}\label{inventory-of-phases-of-the-ism}
\vspace*{-6pt}
Molecular gas is the immediate fuel for star formation, and its spatial distribution, kinematics, and excitation determine how efficiently gas collapses and forms stars. Because H$_2$ lacks a dipole moment, it cannot be detected in dipolar rotational transitions. The most abundant molecule after H$_2$, CO, provides a suitable proxy due to its brightness and well-understood rotational ladder. With multiple rotational transitions (J = 1 $\rightarrow$ 0, up to 4 $\rightarrow$ 3), and isotopologues (e.g., $^{13}$CO, C$^{18}$O) it is possible to disentangle optical depth and constrain density and temperature. At low metallicity, reduced dust shielding and elevated radiation fields expand the CO-dark H$_2$ envelope, partially decoupling CO brightness from total H$_2$ mass, which in turn requires recalibration of the CO-to-H$_2$ conversion factor $X_{\rm CO}$. Contiguous maps of CO, [C\,I], and joint dust analysis in the Clouds provide robust calibrations of $X_{\rm CO}$ and reveal how molecular gas survives and evolves under conditions akin to the early universe \cite{Chevance_20_CO,Meixner_13_Herschel,Gordon2014}.
\vspace*{-12pt}
\paragraph{Chemistry, isotopologues, and CO-dark H$_2$}
A comprehensive survey should target CO and its isotopologues ($^{13}$CO, C$^{18}$O) in different rotational transitions, and [C\,I], which becomes a key tracer where CO is photodissociated but carbon remains neutral. Ratios of CO transitions and isotopologues probe optical depth and excitation, enabling estimates of column density and volume density. In tandem with dust-based gas mass estimators, these measurements allow direct quantification of the CO-dark fraction and its dependence on $A_V$, radiation field strength, and ambient pressure, and provides a measure to the still elusive dust-to-gas ratio. The Magellanic Clouds thus offer the opportunity to build physically motivated $X_{\rm CO}$ calibrations that can be applied to the low metallicity systems at high redshifts.
\vspace*{-12pt}
\paragraph{Kinematics, turbulence, and feedback}
Velocity-resolved mapping across entire galaxies reveals how turbulence is driven and dissipated, how filaments assemble, and how feedback from massive stars sculpts the ISM. Linewidth--size relations and spatial power spectra derived from contiguous datacubes diagnose energy injection scales and the role of shear and tidal forces. In the LMC and SMC, where shells, supershells, and tidal features are ubiquitous, widefield coverage is essential to follow velocity-coherent structures and to connect local cloud evolution to global dynamics.

\vspace*{-12pt}
\subsection{Current Status and Gaps}
\vspace*{-6pt}
At present, the only contiguous CO(1--0) maps at galactic scales are from the NANTEN telescope, which surveyed the LMC and SMC with beams of $\sim$2.6$'$ at 115\,GHz \cite{Fukui_08_Second}. These maps identify cloud-scale ``islands'' of emission, but the coarse resolution leads to significant beam dilution and blurs substructure critical for understanding cloud physics and turbulent energy cascades. A number of the brighter regions have been followed up at higher spatial resolution and in higher rotational transitions (see Sect.~\ref{scientific-context-and-motivation}), but unbiased surveys in higher transitions have only been attempted for sections of the SMC (\cite{Tokuda_21_Unbiased}; Jameson et al.~in prep.). Full surveys of the LMC and SMC in CO(4-3) and CI(1-0) are being planned by the GEco project \cite{Simon_23_Cycling} with the CHAI 64-pixel receiver on FYST, which will be located at 5500m on the slopes of Cerro Chajnantor, 500m above the plateau.
{\bf Thus, a statistically representative census of the molecular reservoir and its spatial distribution at high spatial resolution is still lacking; obtaining such a census would enable us to construct the demographics of cloud masses and sizes, linewidth--size relations, and cloud lifetimes across the full environmental dynamic range present in the Magellanic Clouds.}
\vspace*{-12pt}
\section{What facility would be needed?}
\vspace*{-6pt}
Interferometers deliver superb resolution on compact scales but have small instantaneous fields-of-view and filter large-scale emission, making them unsuitable for assembling fully sampled, degree-scale spectral line cubes across the large fields-of-view required to map the LMC and SMC. Current single-dish submillimeter/millimeter telescopes in the South, including APEX and AMT, typically have apertures $\leq$12\,m and either single or few-pixel heterodyne receivers, leading to modest mapping speeds and large beams (tens of arcseconds to arcminutes) in the (sub-)millimeter wavelength regime. A large-aperture ($\sim$50-m class) single dish with an instantaneous field-of-view ($\gtrsim$2$^\circ$) and multi-pixel heterodyne spectrometers (tens to hundreds of beams), operated from a high, dry site such as the 5000-m high Chajnantor plateau, is required to deliver fast, sensitive, degree-scale spectral mapping with total-power fidelity and a spatial resolution ranging from 1.5$''$ at 950 GHz (sub-pc resolution in the LMC/SMC) to 10$''$ at 30 GHz. The AtLAST concept meets these criteria \cite{Mroczkowski_25_conceptual}. 

A recent white paper \cite{Liu_24_Atacama} on Nearby Galaxies emphasises in particular the unique opportunity for AtLAST to map the molecular content of the Magellanic Clouds. 
\
Relative to current 12-m single-dish facilities in the South, such as APEX, a 50-m facility with a megapixel spectrometer improves the instantaneous mapping speed by $>10^3$ and up to 10$^5$ times when accounting for both aperture and pixel count, and reduces the beam size by $\sim$4$\times$, mitigating beam dilution and reducing integration times in diffuse regions. A similar increase in mapping speed can also be achieved with respect to ALMA, while AtLAST's continuum sensitivity is comparable to that expected from the entire ALMA array after the Wideband Sensitivity Upgrade.   
Additionally, the 2$^\circ$ instantaneous field-of-view enables a high sensitivity to large scale structures, allowing to recover faint, extended low surface brightness structures that are missed by interferometers. 

\bibliography{magellanic}

\end{document}